# Incremental learning with online SVMs on LiDAR sensory data

Le Dinh Van Khoa
The University of Nottingham, Malaysia Campus
Jalan Broga, 43500 Semenyih
Selangor Darul Ehsan, Malaysia
Telephone number, +60 11 2372 1193
hcxdl1@nottingham.edu.my

Zhiyuan Chen
The University of Nottingham, Malaysia Campus
Jalan Broga, 43500 Semenyih
Selangor Darul Ehsan, Malaysia
Telephone number, +6 (03) 8924 8141
Zhiyuan.Chen@nottingham.edu.my

## ABSTRACT

The pipelines transmission system is one of the growing aspects, which has existed for a long time in the energy industry. The cost of in-pipe exploration for maintaining service always draws lots of attention in this industry. Normally exploration methods (e.g. Magnetic flux leakage and eddy current) will establish the sensors stationary for each pipe milestone or carry sensors to travel inside the pipe. It makes the maintenance process very difficult due to the massive amount of sensors. One of the solutions is to implement machine learning techniques for the analysis of sensory data. Although SVMs can resolve this issue with kernel trick, the problem is that computing the kernel depends on the data size too. It is because the process can be exaggerated quickly if the number of support vectors becomes really large. Particularly LiDAR spins with an extremely rapid rate and the flow of input data might eventually lead to massive expansion. In our proposed approach, each sample is learned in an instant way and the supported kernel is computed simultaneously. In this research, incremental learning approach with online support vector machines (SVMs) is presented, which aims to deal with LiDAR sensory data only.

## 1. INTRODUCTION

The LiDAR device has been developed and implemented in various domains, including 3D-image reconstruction, self-driving machine, geographical remote sensing [1]–[4]. By recording the distances measurements, its data becomes effectively to visualize the environment structure. Exploiting the intensive beam of light, the LiDAR captured the distances from its source to the bouncing surface in very high accuracy. The supremacy traveling speed and eco-friendly favor LiDAR solution over other bouncing distance approaches such as ultrasonic. As exploiting the light traveling speed, LiDAR sensor returns abundant data in a rapid time [5]. Consequently, analyzing LiDAR sensory data leads to a large-scale problem and a solution that utilizes the storage capacity must be adopted. Therefore, a large-scale solution problem that makes used of the structural information becomes the top priority in LiDAR sensory data.

Recently, support vector machine has been reported to be the most suitable solution in many signals processing problem [6]. In terms of structural risk minimization, it has been considered as state-of-the-art due to its superior performance and solid mathematical background [7]. The solution to optimizing the defined problem in SVM can be approached in many directions. Extensive work from J.Shawe-Taylor has categorized SVM optimization methodologies into 7 groups namely: interior point; chunking, sequential minimal optimization (SMO); coordinate descent; active set method; Newton's method; stochastic sub-gradient; and cutting plane [8]. In many problems, the selection of an appropriate optimization solution affects the training process. It is advised that interior point algorithms are reliable and accurate to handle problems having the size of thousands of samples. For large-scale problems, in order to handle the model capacity efficiency, the sparsity of dual variables or compact representation must be adopted in the solution [9].

In general, the common approaches for optimizing the SVM problems consider solving the quadratic programming (QP) problem in its dual forms. In interior points approaches, the Cholesky decomposition processed has been evaluated. The approaches simplified the objective function and the constraints into solvable components with linear algebra techniques. However, this technique involves calculating a matrix scaled by the number of training samples. The result from this model escalates heavily the resource capacity and exhausted time-consuming for training. Hence, the interior point algorithm only efficient with small scale problem. In SMO, the method exploits the equity constraints and the property of chunking approaches. The solution is preserved if the columns and rows correspond to the zero entries coefficient are removed. The SMO tries to solve the sub-optimal problem with adjusting a pair of Lagrange coefficient in each sequence. The analytical bounding box constraint saves its process from the QP numerical calculation [9]. Hence, the performance of SMO has been reported to be approximately hundreds of times faster in some problem as opposed to the interior point methods [10].

Since applying box constraint, the convergence rate of SMO approach heavily depends on the trade-off parameter *C* between regularization and error. Therefore, in the high accuracy constraints, the approach requires more computations in order to reach the critical point [11]. In the problem with thousands of support vectors, instead of examining a pair of vectors, the additional process involves in a larger working set or solving the QP sub-problem are adopted.

However, the classical off-line learning, some refers to batch learning, does not fully examine the whole data given restricted time. The off-line learning mechanism of kernel classifiers also prohibits their models to resume the training. The unprocessed data must be retrained from scratch which is time-consuming and inefficient if the model is required in a time interval. To overcome the drawback, training paradigm in streaming fashion, where data input sequentially, is encouraged.

The follows of this paper is organized with respect to these criteria. In section 2, a list of common online approaches focusing on kernel machine methods, with SVM is the main representative, is presented. In section 3, the experiments by using incremental SVM (ISVM) on LiDAR data are described. The comparison of online and offline-learning on LiDAR data is discussed in section 4. In the section, the advantages and limitations of using online model are clarified. The last section, section 5, summarized the result and potential implementation on other fields.

## 2. RELATED WORK

The most common incremental SVM was introduced by Cauwenberghs & Poggio, which solves online SVM directly by considering the status of active sets [12]. A new sample is inserted to the appropriate set by determining the closest marginal from the sample to each set. The optimal solution is updated by the process of adjusting existing samples to the relevant sets after new insertion. However, the approach is favored as a QP optimization problem rather than solving online classifier problem. Consequently, little of successful practical applications utilizing the approach have been reported.

Due to the distinct description during the optimal process, many extensions have been developed to enhance this approach solution. Almost at the same time, Ma and Martin recognized the rules of removing or adding samples to appropriate set for unlearning process [13][14]. The defined rules justify the searching conditions and direct its path to the optimal solution efficiently. The method implements a bookkeeping procedure that records the actions of transferring samples among sets when a new sample is added to the training set. Practical implementation on time-series forecasting with cross-validation mechanism denotes the efficient of bookkeeping as opposed to the traditional batch learning.

Similarly, the work of Martin extends the incremental SVM in classification tasks to function approximation. The searching rule relies on the Kahn Kuhn-Tucker (KKT) conditions and adjusts their multiplier β. The modification is applied with respect to reserve the constraints conditions of remaining data. In general, the training sequence consists of 3 main processes, incrementally add new data to the training set, remove data from the support set, and update target values for existing data. A comprehensive study indicates the main drawbacks of incremental SVM at the increasing of computation in quadratic time. The complexity heavily depends on the balance of memory access and arithmetic operation [15]. Therefore, implementation of ISVM is not favored as other powerful batch learning packages like LibSVM or SVM light.

### 2.1 Incremental SVM

The proposed solution of Laskov et al. directly restructures the accounting storage and reorganizes the computations [15]. The solution is considered as a lossless model since it maintains all of the observed data and arranges them in the appropriate sets. The approach exploits the KKT conditions which are defined

$$g_i = \alpha K_i + \mu y_i - 1 \begin{cases} \geq 0 & if\ \alpha_i = 0\ \ (remainder\ set - r) \\ = 0 & if\ 0 < \alpha_i < C\ (support\ set - s) \\ \leq 0 & if\ \alpha_i = 0\ \ \ \ \ \ (error\ set - e) \end{cases}$$

$$y^T \alpha = 0$$

When new streaming data c is input, the Lagrangian coefficients must be adjusted to satisfy the constraints. Instead of solving the minimax problem of SVM batch learning, ISVM considers minimizing the loss of new sample with previously observed data. To enhance the efficiency of computation, a compact matrix Q denotes the kernel representation of support sets and their sign has been introduced. The compact representation of changed variations are described as

$$\beta = -\begin{bmatrix} 0 & \alpha_s^T \\ \alpha_s & K_{ss} \end{bmatrix}^{-1} \begin{bmatrix} \alpha_c \\ K_{cs}^T \end{bmatrix} = -Q^{-1}\vec{\eta}$$

$$\gamma = \begin{bmatrix} y_c & K_{cs} \\ y_r & K_{rs} \end{bmatrix}\beta + \begin{bmatrix} K_{cc} \\ K_{cr}^T \end{bmatrix}$$

where β indicates the sensitivity of observed data from the support set with respects to the new sample c; and γ indicates the sensitivity of margin from the remainder set with respects to the new sample c. The largest possible increment of the new sample is determined by a bookkeeping procedure. The procedure is account for the changing structure when a sample reaches its set variation. 4 possible cases of constraints violation have been analyzed: a support coefficient reaches its bounding constraints; a remainder sample shifts to the margin when $g_i$ closes to 0; the new sample belongs to support set which requires updating the other coefficient, and the new sample coefficient reaches upper bound constraint. The moving sample that yields the minimum variation is transferred to the relevant set. Once the new sample is allocated to the correct set, the inverse matrix Q is expanded with an additional zero row and column, and its updated result is obtained by matrix multiplication.

$$\tilde{Q} = \begin{bmatrix} Q^{-1} & \eta_k \\ \eta_k^T & K_{kk} \end{bmatrix}^{-1} = \begin{bmatrix} Q & 0 \\ 0 & 0 \end{bmatrix} + \frac{1}{\kappa}\begin{bmatrix} \beta_k \\ 1 \end{bmatrix}[\beta_k^T\ \ 1]$$

in which $\kappa = K_{kk} - \eta_k^T Q \eta_k$

Hence, the operation time for updating and removing is quadratic in the size of $Q$ [15]. Although ISVM describes exactly the process of online-learning, the computation quickly escalates with the number of learned data. According to the learning mechanism, ISVM has to record the entire data and the belonging status. The operation time of ISVM boosts rapidly at the very first samples and degrade linearly in later iterations.

### 2.2 Online LASVM

LASVM is a semi-online training mechanism that is also applicable to other kernel classifiers [16]. The approach solves the large margin classifier problem by utilizing the sequential searching direction of SMO. The direction, called $\tau$-violating pair, is determined by moving along a pair of samples $(i,j)$ as long as it expands the margin without violating any constraint.

$$\tau - violating\ pair\ (i,j) \Leftrightarrow \begin{cases} \alpha_i < \max(0, Cy_i) \\ \alpha_j > \min(0, Cy_j) \\ g_i - g_j > \tau \end{cases}$$

The convergence of solution is achieved by alternating two phases of direction search, namely PROCESS, and REPROCESS. In PROCESS phase, a potential vector $(i)$ is considered to be appended into the current kernel. At initial, the new sample is added to the support set. Then, the process identifies its second $\tau$-violating pair $(j)$ from the support set $S$ that has the greatest gradient. The searching directions of existing support vectors are shifted accordingly

$$\lambda = \min\left(\frac{g_i - g_j}{K_{ii} + K_{jj} - 2K_{ij}}, \max(0, Cy_i) - \alpha_i, \alpha_j - \min(0, Cy_j)\right)$$

$$\alpha_i = \alpha_i + \lambda\ ;\ \alpha_j = \alpha_j - \lambda$$

$$g_s = g_s - \lambda(K_{is} - K_{js})\ \ \ \forall\ s \in S$$

On the contrary, instead of preserving all of the observed data in ISVM, LASVM adopts the removal mechanism to manage the storage capacity efficiently. The elimination procedure is achieved in the REPROCESS. This process, first, repeats the searching for $\tau$-*violating pair* $(i, j)$ as in the previous description.

$i = \arg max_{s \in S}$ where ; $\alpha_s < \max(0, Cy_s)$

$j = \arg min_{s \in S}$ where ; $\alpha_s > \min(0, Cy_s)$

Upon completion of the adjustment, any support vectors that exceed the new bounding constraint –defined by the pair $(i, j)$ -are pruned. At the end of elimination, the bias term of decision function and the gradient of $\tau$-*violating pair* $(i, j)$ are recomputed.

$b = \dfrac{g_i + g_j}{2}$

$\delta = g_i - g_j$

LASVM successfully combines online and offline-learning in its processes. In the online iterations, the adding and removing procedure are achieved consequently through PROCESS and REPROCESS. However, to reach a better solution, additional REPROCESS steps must be applied gradually until there is no further $\tau$-*violating pair*, defined as $\delta < \tau$. The finishing step performs as offline-learning since it is achieved after the entire batch has been observed.

In the practical implementation, the online iteration of LASVM is learned through in epochs. An epoch is defined by a sequence of shuffle training example. Running one epoch involve in the computation of online setup. To ensure the accuracy of output model, the finishing step is applied after a predefined number of epochs. Multiple epochs are combined as a stochastic optimization approach [16]. Report from different benchmark dataset indicates the competitive accuracy with common offline SVM in a single training. Moreover, LASVM requires less memory and also dominate common SVM solvers in training time.

## 3. EXPERIMENTS

The machine learning scikit-learn package has been imported to produce LibSVM model for the offline-learning. The reproduced code of ISVM from the Matlab project is also inherited to build the online model. For the semi-online model LASVM, the code has been rewritten from the author's C-version. All of the experiments have been conducted in an Intel 8700 (6 core/ 12 threads/ 4.6 Ghz), 16 Gb RAM workstation.

Real world problem with LiDAR sensory data has been used in the experiment. The dataset is obtained by sending an inspection gauge with the attached LiDAR sensor through the pipeline object. This LiDAR dataset, which measured the distance of signal sending from sensor to the bouncing pipe, consists of 11785 samples (both healthy and defected instances) have been captured when scanning the pipeline. The training and testing sets have been separated by the 3:7 ratio. The samples in the training set are also split 20% for the model validation. To preserve a fair and objective evaluation, all of the hyper-parameters in each model are carefully tuned with Grid Search (GS). The same type of kernel is applied to produce a comparison among models. Firstly, the GV is applied to all 3 approaches to determine the most suitable hyper-parameters. The searching was conducted in a simpler pipe dataset consists of 863 instances. 5-fold cross validation was adopted to verify the most suitable hyper-parameters. The range of each hyper-parameters and its best performance is specified in table 1.

**Table 1. List of hyper-parameters tuning and the best configuration**

| Hyper-parameter | Values | Best | | |
|---|---|---|---|---|
| | | ISVM | LASVM | offline |
| C | *0.1, 0.5, 1, 2, 5, 10, 20, 25, 50, 75, 100, 150, 200* | 100 | 100 | 100 |
| Kernel | *Polynomial, RBF, Sigmoid, Chi-square* | RBF | RBF | RBF |
| γ-kernel | *auto, 0.001, 00.1, 0.1, 1, 10* | auto | auto | 0.001 |
| $\tau$-*violating* | *0.1, 0.05, 0.01, 0.001* | _ | 0.01 | _ |

During the hyper-parameter tuning, we recognized that by using the first 517 samples and trained them with the online fashion; the performance on test set already reached the accuracy of 92%. Although this result is not as good as offline performance, whereas the offline result reached more than 98%, the online approach reduced the workload of offline-learning. In particular, the average time for learning 1 fold (517-samples) is approximate 90s while the offline approach required more than 380s to complete 1 fold. The process of updating kernel matrix when new sample belongs to support set is also simplified in online-learning by appending an additional row and column to the kernel matrix of the learned instances. Besides, the online trained model was saved whenever adding a new sample. As a result, it can be used immediately to evaluate the performance on the validation set.

In term of the kernel selection, both offline and online method acknowledged RBF with rather small γ as producing the most relevant output. This result simplifies the comparison procedure by selecting the same set of best performance hyper-parameters to train in the final dataset. Each model has been conducted 10 times to achieve all of the results accuracy and the operational time. The mean value of these results is computed in the final comparison. All of the evaluation metrics are also derived by using scikit-learn package.

## 4. RESULTS & DISCUSSION

In the first few hundreds of samples, ISVM solves the active set problem slightly better than LASVM and offline method. For a small dataset, the separation hyperplane spans more freely and faster in a given Hilbert space. Since a few vectors violate the KKT constraints in the early training stage, the learning process of ISVM mainly focuses on adding new sample and updates the relevant kernel. On the contrary, offline-learning and LASVM (partially) have to compute the $\tau$-violating pair to reorganize the support set.

However, as the number of instances increases to thousands, the operation time of LASVM quickly surpasses ISVM. The complexity of ISVM adjusting rules requires heavy computation that restrains the training speed. Since the separation hyperplane has been well-defined, any misclassification from the new data results in a sequence of reallocating vectors. In the worst case, most of the vectors have to be reallocated, which scales the computation time with the number of learned vectors.

Unlike ISVM, most of the expensive computation of LASVM only occurs in the offline phase. During the online phase, the computation is extremely fast as only alternating the adding and one-shot removal verification. The misclassifications are compensated on completing a predefined number of epoch or the end of training samples. In the experiment, an epoch is set as training a batch of each 200 samples, and after 5 epochs, the finishing step is executed. In addition, the reorganizing stage of LASVM also involves in permanently remove the non-contribution vectors. Therefore, the adjustment in LASVM computes fewer vectors, and, hence, shortens the operation time.

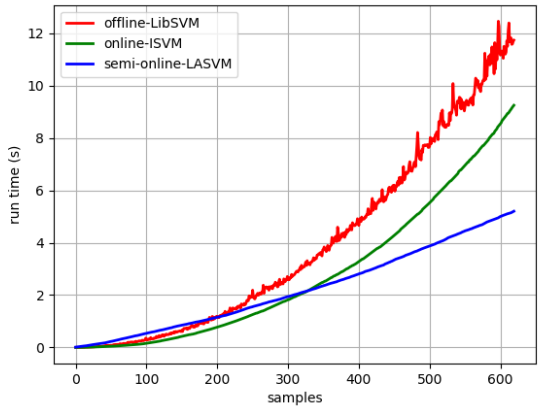

**Figure 1 The comparison of learning time with respect to the number of trained data**

To boost up the training speed, LASVM forfeits the converging step during the online process. Particularly, its main objective has been simplified to add a new sample. Only one-time verification of the suitability of the sample in that set is executed. As a result, in the accuracy plot, the performance usually drops heavily at the middle of a new epoch, and then get recover before reaching the fittest point at the end of the finishing step. Despite effectively achieve saddle point through the offline process, it remains as the drawback of LASVM. Thus, LASVM cannot be recognized as a fully standalone online approach.

**Table 2. Table captions should be placed above the table**

| Model | Accuracy | Log-loss | ROC-AUC | F1 |
|---|---|---|---|---|
| ================*Validation Set*================ | | | | |
| LibSVM | 99.58 | 0.1466 | 0.9963 | 0.9951 |
| ISVM | 98.73 | 0.4397 | 0.9853 | 0.9851 |
| LASVM | 97.88 | 0.7328 | 0.9813 | 0.9761 |
| ==================*Test Set*=================== | | | | |
| LibSVM | 99.43 | 0.1967 | 0.9949 | 0.9936 |
| ISVM | 98.52 | 0.5107 | 0.9832 | 0.9829 |
| LASVM | 98.13 | 0.6447 | 0.9834 | 0.9792 |

As to standing side by side with the offline method, the accuracy derived from both LASVM and ISVM are competitive. Figure 2 indicates that all of the approaches learned the best generalization bound and converge after examining 2000 samples. In comparison with ISVM, LASVM performs more stable as it firmly achieves a better solution when new data arrives. Especially, the learning curve of LASVM is closed to the behavior of the offline-learning. This reflects the fact that LASVM implies resemble chunking technique with offline-learning. Although ISVM did not improve steadily, it dominates LASVM on some occasion, as well as in average performance.

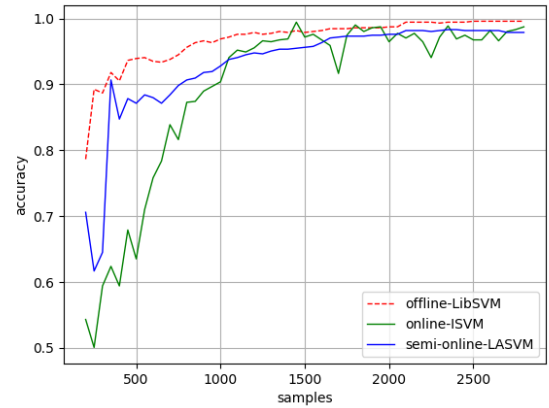

**Figure 2 Validation accuracy with respect to the number of training samples**

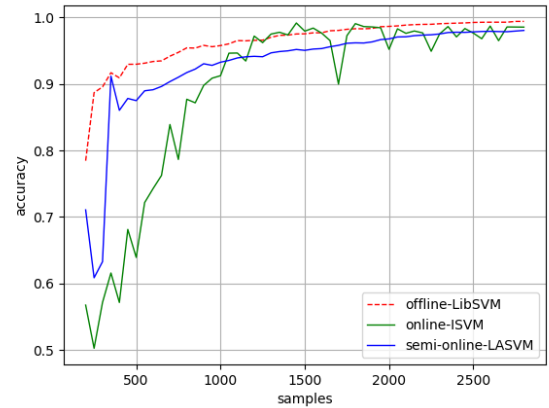

**Figure 3 Testing accuracy with respect to the number of training samples**

The slow convergence denotes that ISVM solves the problem without having the overall information. For any model, the main objective is to define a separation hyperplane with the least misclassifications. However, if the sequence of examining samples (the identity and samples presented order) is informed in offline-learning, in online-learning, the classifier only received the set of samples. [17]. Hence, ISVM is challenged with 2 types of uncertainties. The first is to identify the most appropriate set allocation and the relevant coefficient for the consistent of learned data. In addition, ISVM has to recognize which instances would be challenged in the future. While the first problem is common to any SVM classifier, the second challenge is specified for incremental learning.

## 5. CONCLUSION

The use of online-learning has been indicated to be essential for large scale problem. System, which utilizes LiDAR data, has the tendency to expand its data size tremendously over time. As a result, an online model must be considered to tackle the problem appropriately. In this work, online learning approaches with SVM classifier, which are ISVM and LASVM, have been examined. Performance has indicated the competitive test error of these 2 models with common offline SVM solver. Given the comparable accuracy, ISVM and LASVM gain its benefit from offline SVM by solving the problem faster. In addition, the models can be extracted during the training process and resume its training without recomputed from the beginning.

## 6. ACKNOWLEDGMENTS

Our thanks to MOSTI grant from the University of Nottingham for supporting us in this project

Project code 01-02-12-SF0360